\documentclass[a4paper,11pt]{article}
\usepackage{pos}
\usepackage{braket}

\usepackage{amsmath, amsfonts, hyperref,braket,amssymb, xcolor, tikz}
\usetikzlibrary{decorations.markings}

\title{Towards the phase diagram of fermions coupled with $SO(3)$ quantum links in $(2+1)-$D}
\ShortTitle{Phase diagram of $(2+1)-$D $SO(3)$ QLM with matter}

\author*[a]{Graham Van Goffrier}
\author[b]{Debasish Banerjee}
\author[a]{Bipasha Chakraborty}
\author[c]{Emilie Huffman}
\author[b]{Sandip Maiti}

\affiliation[a]{Department of Physics and Astronomy, University of Southampton,\\
 University Road, Southampton SO17 1BJ, UK}

\affiliation[b]{Theory Division, Saha Institute of Nuclear Physics, 
1/AF Bidhannagar, Kolkata 700064, India}
\affiliation[b]{Homi Bhabha National Institute, Training School Complex, 
Anushaktinagar, Mumbai 400094, India}

\affiliation[c]{
Perimeter Institute for Theoretical Physics,\\
31 Caroline St N, Waterloo, ON N2L 2Y5, Canada}

\emailAdd{gwvg1e23@soton.ac.uk}
\emailAdd{debasish.banerjee@saha.ac.in}
\emailAdd{b.chakraborty@soton.ac.uk}
\emailAdd{ehuffman@perimeterinstitute.ca}
\emailAdd{sandip.maiti@saha.ac.in}

\abstract{Quantum link models (QLMs) are generalizations of Wilson's
lattice gauge theory formulated with finite-dimensional link Hilbert 
spaces. In certain cases, the non-Abelian Gauss Law constraint can be
exactly solved, and the gauge invariant subspace embedded onto local 
spin Hamiltonians for efficient quantum simulation. In $(1+1)d$ previous
studies of the $SO(3)$ QLM coupled to adjoint fermionic matter have been 
shown to reflect key properties of QCD and nuclear physics, including 
distinct confining/deconfining phases and hadronic bound states. We extend
the model to $(2+1)d$ dimensions for the first time, and report on our 
initial results. We review the construction of gauge-invariant state space 
for the proposed models, and study the single-plaquette ground state via 
exact-diagonalisation. We provide indications of a rich phase diagram which
shows both spontaneous and explicit chiral symmetry breaking, confinement, 
and distinct magnetic phases characterised by different plaquette expectation values.}

\FullConference{%
  The 41st International Symposium on Lattice Field Theory (Lattice 2024) \\
  July 28th - August 3rd, 2024 \\
  University of Liverpool, Liverpool, UK
}

\begin{document}
\maketitle

\newcommand{\bc}[1]{\textcolor{orange}{(#1)}}
\newcommand{\db}[1]{\textcolor{green}{(#1)}}
\newcommand{\eh}[1]{\textcolor{blue}{(#1)}}
\newcommand{\gvg}[1]{\textcolor{red}{(#1)}}
\newcommand{\sm}[1]{\textcolor{purple}{(#1)}}

\section{Introduction}

Lattice gauge theories (LGTs) have proven in recent decades to be an 
indispensable tool for extracting quantitative predictions from the 
Standard Model, especially non-perturbative contributions stemming 
from quantum chromodynamics (QCD). In combination with renormalized 
perturbative calculations, lattice QCD (LQCD) has allowed numerous 
measured observables to be compared with high-quality theoretical values.

As a computational framework, LQCD has typically relied upon the 
Monte Carlo sampling of path-integral formulated in Euclidean time, 
to avoid the sign problem. While static, ground state, and thermodynamic
properties can be computed very well in this approach, studies of 
real-time QCD dynamics and properties of QCD matter at finite baryon 
density suffer from a severe sign problem. 
It is widely hoped \cite{wiese2014towards} that quantum computation 
may allow for the simulation of LQCD and other many-body quantum systems 
in the long-term without exponentially-scaling resources.

In this work we turn to quantum link models (QLMs) \cite{wiese2022quantum} 
as a framework that is especially suited to quantum computation 
approaches. In contrast to the Wilsonian formulation, the quantum link
formulation provides an additional parameter: the representation of link
operators, which controls the Hilbert space of local gauge fields and has an exact local gauge invariance. 
Each such choice of representation leads to a valid theory in its own right with that same gauge invariance.
There is thus no need for a gauge-symmetry breaking truncation of the infinite-dimensional Hilbert 
space needed for each gauge link in the Wilson gauge formulation. QLMs have 
been studied with increasing interest by several groups, including 
through classical simulations employing exact-diagonalisation (ED) 
and tensor networks (TN) 
\cite{halimeh2022achieving,zache2022toward,huang2019dynamical}, and 
on real quantum hardware 
\cite{huffman2022toward,banerjee2022nematic,osborne2305spin}. Most often, 
Abelian $U(1)$ QLMs have been investigated, with and without
matter, but only a few results exist for the physics of $SU(2)$ and 
$SO(3)$ gauge theories 
\cite{boegli2014thesis, rico2018so, maiti2024spontaneous}.

We choose to focus on the $SO(3)$ QLM because of qualitative properties 
it is known to share with QCD, including fermionic baryon bound states, 
and spontaneous chiral symmetry breaking (in $(1+1)d$). 
Refs.~\cite{boegli2014thesis} and \cite{rico2018so} were the first to 
demonstrate these properties with numerical simulations in $(1+1)d$, 
including with dynamical matter fermions. More recently, a subset of us 
\cite{maiti2024spontaneous} have studied the matter-free $SO(3)$ 
QLM in $(2+1)d$ using quantum algorithms to demonstrate the spontaneous
symmetry breaking in that theory.

Here we present the first ED results obtained for the $SO(3)$ QLM in 
$(2+1)d$ including dynamical matter fermions, and show that even on a 
single plaquette, the model demonstrates both explicit and spontaneous 
chiral symmetry breaking, as well as other intriguing properties such 
as distinct magnetic phases. In Section 2, we define the $SO(3)$ QLM 
and show how to solve the Gauss's Law constraint in $(2+1)d$, arriving 
at a fully gauge-invariant Hilbert space. In Section 3 we overview our 
ED techniques and the symmetry-driven speedups which enable such a study, 
and in Section 4 we deliver our numerical results, before discussing 
implications and continuing work in Section 5.

\section{$SO(3)$ Quantum Link Models with Fermions}

QLMs succeed in exactly preserving local gauge invariance by modifying 
the commutation relations between the quantum link fields, which do not
affect the gauge invariance. Consequently, the operators are not unitary.
Explicit representations of the link operators and their canonical momenta
(the electric fields) can obtained via an appropriate embedding algebra. 
Following the notation of \cite{rico2018so}, we review here the gauge 
and fermion operators of our construction, implement the $so(6)$ embedding 
algebra, and then solve Gauss's Law.

\subsection{Operators and Gauge-Invariant States}

The complete $SO(3)$ operator algebra on each site is:
\begin{align}
   [L^a, L^b] = 2 i \varepsilon^{abc} L^c &, [R^a, R^b] = 2 i \varepsilon^{abc} R^c, \nonumber\\
   [L^a, O^{bd}] = 2 i \varepsilon^{abc} O^{cd} &, [R^a, R^{bd}] = - 2 i \varepsilon^{abc} O^{cd}, \nonumber\\
   [O^{ab}, O^{cd}] = 2 i \delta^{ac} &\varepsilon^{ebd} R^e + 2 i \delta^{bd} \varepsilon^{eac} L^e, 
\end{align}
where $L^a$, $R^a$ and $O^{ab}$ are the $SO(3)$ matrix-valued left/right 
electric fields and gauge fields, respectively, with colour indices 
$a,b \in 1,2,3$ while spacetime labels are suppressed. We include staggered 
fermionic matter fields $\psi^a$ which transform in the adjoint representation 
of $SO(3)$. This has the important consequence that the colour of the 
fermion can be screened by the colour of a gauge field in order to form
a hadronic state, which can be fermionic in nature, as in QCD. In contrast, 
while considering matter in the fundamental representation, one only obtains 
bosonic hadronic states. 

The generators of gauge transformations at each site may then be written 
\cite{rico2018so}:
\begin{equation}
    G^a = - 2 i \psi^{b\dagger} \varepsilon^{abc} \psi^c 
    + \sum_k \left( L^a_{x,+\hat{k}} + R^a_{x,-\hat{k}} \right),
\end{equation}
such that gauge-invariant states must satisfy $G^a \ket{\Psi} = 0$; this 
is Gauss's Law. It is also standard practice to encode the above operator 
algebra by means of an $\frak{so}(6)$ embedding algebra composed of spin 
bilinears:
\begin{equation}
    O^{ab}_{x,x+\hat{k}} = \sigma^a_{x,+\hat{k}} \otimes \sigma^b_{x+\hat{k},-\hat{k}}, \quad L^a_{x,+\hat{k}} = \sigma^a_{x,+\hat{k}} \otimes \mathbb{I},
    \quad R^a_{x,-\hat{k}} = \mathbb{I} \otimes \sigma^a_{x,-\hat{k}}.
\end{equation}
When we write down gauge-field states, we can then use the notation 
$\ket{\uparrow},\ket{\downarrow}$ for spin states, as well as 
the convention $S^a = \frac{1}{2} \sigma^a$ for spin operators.

Schematically, each $\ket{\Psi}$ at a site is a product of gauge-field and fermion 
states, i.e. $\ket{\Psi} = \ket{\chi}_g \ket{\psi}_f$, where on a $(2+1)d$ 
square lattice $\ket{\chi}_g$ will receive contributions from four links. 
A gauge-invariant singlet $\ket{\Psi}$ may therefore be built either from both 
singlet gauge and fermion components, or from ``matching'' 
triplet gauge and fermion components.

The former case is more trivial. $\ket{\psi}_f$ will be a gauge-invariant singlet 
only in the case of fermionic zero occupation $\ket{0}_f$ or maximum occupation 
$\ket{3}_f$. As shown in \cite{rico2018so}, there are two pure-gauge singlets in 
$(2+1)d$, which we denote $\ket{\chi_1}_g$ and $\ket{\chi_2}_g$. Taking 
all products of these, we have identified four singlet-singlet 
gauge-invariant states:
\begin{align}
    \ket{\Psi_1} &= \ket{\chi_1}_g \ket{0}_f, \quad \ket{\Psi_2} = \ket{\chi_2}_g \ket{0}_f, \quad \ket{\Psi_9} = \ket{\chi_1}_g \ket{3}_f , \quad \ket{\Psi_{10}} = \ket{\chi_2}_g \ket{3}_f, \nonumber\\
    \ket{\chi_1}_g &= \frac{1}{2} \left( \ket{\uparrow \downarrow \uparrow \downarrow} - \ket{\uparrow \downarrow \downarrow \uparrow} - \ket{ \downarrow \uparrow \uparrow \downarrow} + \ket{\downarrow \uparrow \downarrow \uparrow} \right), \nonumber \\
    \ket{\chi_2}_g &= \frac{1}{\sqrt{12}} \left( 2 \ket{\uparrow \uparrow \downarrow \downarrow} + 2 \ket{\downarrow \downarrow \uparrow \uparrow} - \ket{\uparrow \downarrow \uparrow \downarrow} - \ket{\uparrow \downarrow \downarrow \uparrow} - \ket{ \downarrow \uparrow \uparrow \downarrow} - \ket{\downarrow \uparrow \downarrow \uparrow} \right),
\end{align}
where the spin (gauge-field) basis is ordered around the vertex, e.g. clockwise.

The latter case requires more construction. Pure-gauge triplet states 
may be obtained by constructing products of a singlet spin-pair on two 
links and a triplet spin-pair on the other two links \cite{boegli2014thesis}, and then 
applying spin-raising operators. Nine independent states result, denoted 
as $\{ \ket{\chi_3}_g, ..., \ket{\chi_{11}}_g \}$. These may be combined 
with single-occupation fermion states $\ket{1}_f$ via the ansatz:
\begin{equation}
    \ket{\Psi} = \left[ \frac{1}{2} (S^-\ket{\chi_{S_z = 0}}_g - S^+ \ket{\chi_{S_z = 0}}_g ) \psi^{1\dagger} + \frac{i}{2} (S^-\ket{\chi_{S_z = 0}}_g + S^+ \ket{\chi_{S_z = 0}}_g ) \psi^{2\dagger} + \ket{\chi_{S_z = 0}}_g \psi^{3\dagger}  \right] \ket{0}_f
\end{equation}
where $\ket{\chi_{S_z = 0}}_g$ may be any of $\ket{\chi_6}_g$, $\ket{\chi_7}_g$,
$\ket{\chi_8}_g$. This ansatz satisfies $G^a \ket{\Psi} = 0$, as does 
the obvious generalisation to double-occupation fermion states. Therefore, 
we have identified six triplet-triplet gauge-invariant states, for a total 
dimensionality of ten gauge-invariant states per site.


\subsection{Action of Gauge-Invariant Operators}

\begin{figure}[h!]
    \centering
    \includegraphics[width=1.0\linewidth]{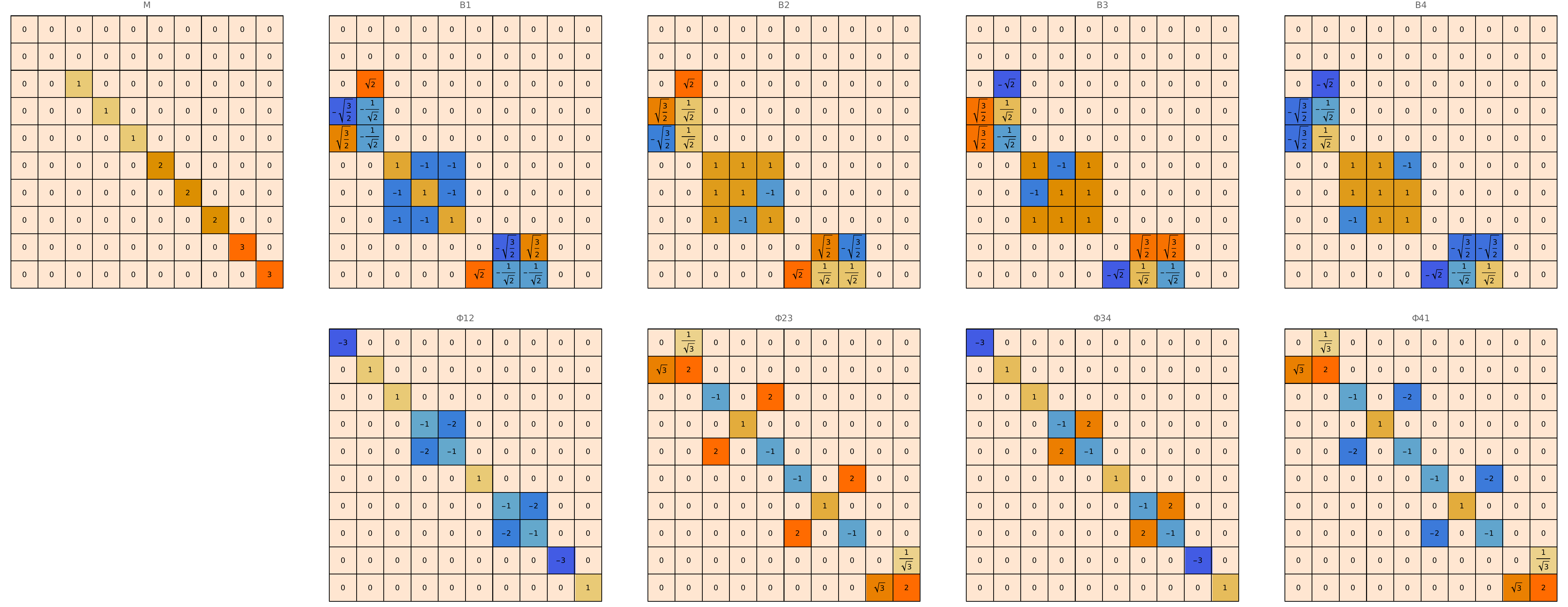}
    \caption{Gauge-invariant operators acting at each site, written 
    in the $\ket{\chi_i}$ basis.}
    \label{fig:matrices}
\end{figure}

The only gauge-invariant operators for the above state space are composed 
of either two fermion-operators, two gauge-operators, or one of each:
\begin{align}
    M = \sum_{a=1}^3 \psi^{a\dagger} \psi^a &, 
    \quad \Phi_{ij} = \sum_{a=1}^3 S_i^a S_j^a \nonumber\\
    B^+_i = \sum_{a=1}^3 2 \psi^{a\dagger} S_i^a &, 
    \quad B^-_i = \sum_{a=1}^3 2 \psi^a S_i^a 
\end{align}
Applying these operators to the gauge-invariant basis defined above, 
we arrive at matrix representations convenient for ED, which are 
illustrated in Figure \ref{fig:matrices}. $M$ is easily seen to count 
fermion occupation number, while $B_i^{\pm}$ are generalised 
raising/lowering operators which mix the gauge degrees of freedom, 
and the $\Phi_{ij}$ are purely gauge-mixing.

\begin{figure}[h!]
  \centering
  \begin{minipage}[b]{0.5\textwidth}
    \begin{center}
         \begin{tikzpicture}[arrowmark/.style 2 args={decoration={markings,mark=at position #1 with \arrow{#2}}}]
         \draw[solid] (4.2,2) -- (1.8,2)node[below] {};
         \draw[solid] (4.2,4) -- (1.8,4)node[above] {};
         \draw[solid] (2,4.2) -- (2,1.8)node[below] {};
         \draw[solid] (4,4.2) -- (4,1.8)node[below] {};
         \draw[ultra thick]  (2,2) rectangle (4,4);
         \node at (2.35,3.7) {$\Phi_{34}$};
         \node at (2.35,2.3) {$\Phi_{23}$};
         \node at (3.65,2.3) {$\Phi_{12}$};
         \node at (3.65,3.7) {$\Phi_{41}$};
         \node at (3,1.7) {$B^+_3 \otimes B^-_1$};
         \node at (3,4.3) {$B^+_3 \otimes B^-_1$};
         \node at (1.2,3) {$B^+_2 \otimes B^-_4$};
         \node at (4.8,3) {$B^+_2 \otimes B^-_4$};
         \node at (1.7,1.7) {$M$};
         \node at (1.7,4.3) {$M$};
         \node at (4.3,4.3) {$M$};
         \node at (4.3,1.7) {$M$};
         \end{tikzpicture}
    \end{center}
    \caption{Operator schematic for a single-plaquette lattice.}
    \label{fig:schematic}
  \end{minipage}
  \hfill
  \begin{minipage}[b]{0.45\textwidth}
    \raisebox{4.4em}{\includegraphics[width=\textwidth]{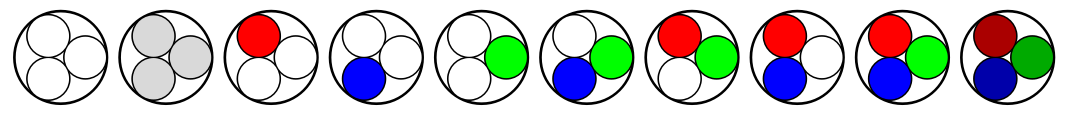}}
    \caption{Symbolic notation for $10$ gauge-invariant states per site.}
    \label{fig:states}
  \end{minipage}
\end{figure}

As shown in \cite{rico2018so}, the Hamiltonian can be re-expressed in terms 
of these operators, effectively pre-diagonalising it in the 
gauge-invariant subspace:
\begin{align}
    H &= - t \sum_{x,\hat{k}} \left[ s_{x,x+\hat{k}} \vec{B}^+_{x,+\hat{k}} \cdot \vec{B}^-_{x+\hat{k},-\hat{k}}  + \text{h.c.} \right] + m \sum_x s_x M_x \nonumber\\
    &- \frac{1}{4 g^2} \sum_{x} \text{Tr}\left[ \Phi_{12} \Phi_{41} \Phi_{34} \Phi_{23} \right] + G \sum_x M_x^2 + V \sum_{x,\hat{k}} M_x M_{x+\hat{k}},
\end{align}
where $s_x = (-1)^{x_1 + x_2}$ and $s_{x,\hat{k}} = (-1)^{x_1 + ... + x_{k-1}}$ 
are alternating signs, spacetime indices on the $\Phi$ operators follow 
a counterclockwise plaquette ordering, and the correspondence between 
axes $\hat{k}$ and indices $i \in \{1,2,3,4\}$ is most easily understood 
from the schematic in Figure \ref{fig:schematic}.

\section{Exact Diagonalisation Methodology}

We perform ED by explicitly constructing the Hamiltonian as a 
sparse matrix whose entries are simple rational functions of 
the physical parameters $\{t,m,g,G,V\}$, then substituting 
chosen values when runs are performed. Each run consists of 
a Lanczos (Arnoldi) iteration procedure to prepare the ground 
state of the Hamiltonian \cite{lanczos1950iteration}. 
Both sparse matrix storage/retrieval 
and Lanczos iteration are accelerated by pre-diagonalising 
the Hamiltonian with respect to its global baryon-number symmetry:
\begin{equation}
    B = \sum_x \left( M_x - \frac{3}{2} \mathbb{I}_{10} \right), \qquad [B, H] = 0.
\end{equation}
Labelling global fermion states of the lattice by a row-major 
ordering of occupation numbers at each site, e.g. 
$\ket{\Psi_9} \otimes \ket{\Psi_3} \otimes \ket{\Psi_8} \otimes 
\ket{\Psi_1} \equiv \ket{3120}$, we observe that the eigenvalue 
of $B$ on a basis state is simply the sum of these occupation 
numbers less $3N/2$, where $N$ is the total number of sites.
Therefore, the $B=0$ sector corresponds to all $N$-part 
partitions (allowing $0$) of $3N/2$, which can be easily computed 
for small $N$. For the single-plaquette $N=4$, there are $44$ such 
states, which expand to $1878$ states once gauge degrees of freedom 
are accounted for, out of $10^4$ total physical states.

Although the baryon number of the ground state is not $0$ for all 
choices of parameters $\{t,m,g,G,V\}$, and we have analytically 
understood its variation with the Fermi couplings $G,V$ in particular, 
we restrict ourselves to the $B=0$ sector for the analysis presented 
here. In the continuum limit, the ground state of $SO(3)$ gauge theory 
is expected to be $B=0$ across all accessible physical parameters, 
where the baryon number symmetry is not broken spontaneously.

\section{Chiral and Magnetic Observables for a Single Plaquette}

A standard gauge-invariant observable in LQCD is the average spatial 
plaquette $\braket{\Phi}$, or in our notation, the expectation value of 
$\text{Tr} [ \Phi \Phi \Phi \Phi ] $. This quantity is known to be a 
good order parameter for bulk thermodynamics, and 
can also be interpreted as an average magnetic field energy. For example, 
in the $(3+1)d$ $Z_2$ lattice gauge theory, $\braket{\Phi}$ exhibits a 
large discontinuity at the first-order phase transition \cite{creutz1980phase,creutz1983generalized}.

Even for a single plaquette, we can expect that $\braket{\Phi}$ may 
behave quite differently at strong- and weak-coupling, and in 
response to four-Fermi couplings. Notably, this observable is 
sensitive to both gauge and fermionic degrees of freedom, and 
is trivially absent from the $(1+1)d$ physics of \cite{rico2018so}.

\begin{figure}[t!]
    \centering
    \includegraphics[width=1.0\linewidth]{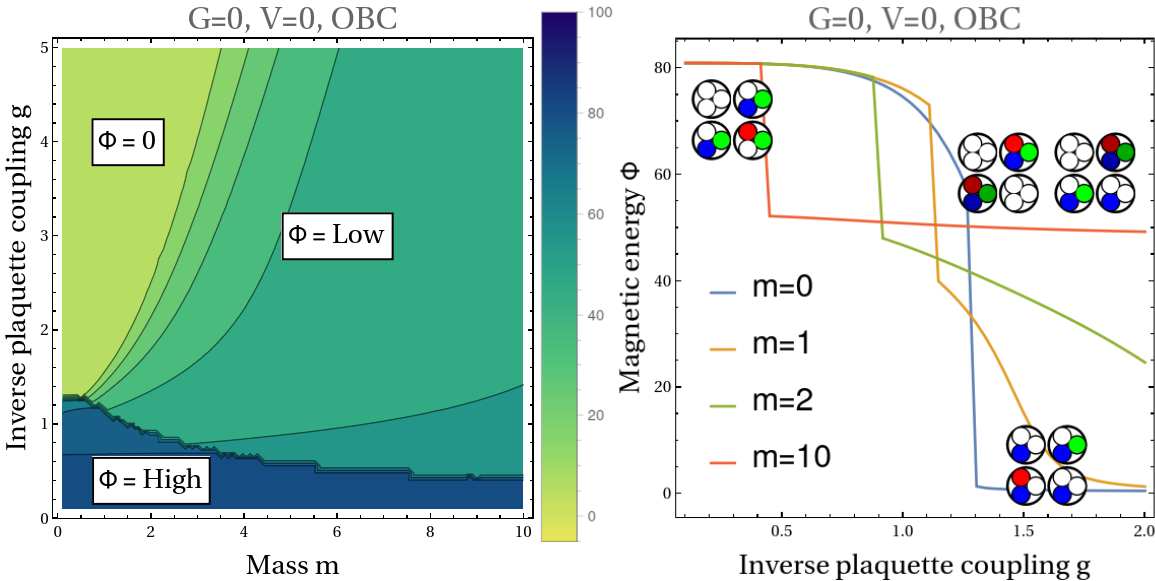}
    \caption{Ground-state plaquette observable $\braket{\Phi}$ for 
    varied fermion mass $m$ and inverse plaquette coupling $g$. }
    \label{fig:results_magnetic}
\end{figure}

Let us first put the four-Fermi couplings to the side, and observe 
via ED on a single-plaquette how $\braket{\Phi}$ varies just with 
inverse plaquette coupling $g$ and fermion mass $m$. The contour 
plot in Figure \ref{fig:results_magnetic} (left) reveals three 
distinct phases: $\braket{\Phi} \sim 0$ (zero-field), $\braket{\Phi} 
\sim 50$ (low), and $\braket{\Phi} \sim 80$ (high). The high-field region occurs only at stronger plaquette couplings (smaller $g$), with the critical coupling 
$g_C$ between zero-field and high-field regions especially sharp in 
the massless limit. As illustrated in \ref{fig:results_magnetic} 
(right), the high-field region becomes narrower for nonzero fermion 
mass $m$. Meanwhile, at weaker plaquette couplings $g$, the introduction 
of mass gradually transitions zero-field behavior to low-field behavior. 
We also remark that increased $\braket{\Phi}$ is associated with a 
confining of fermions, as denoted by state schematics in the Figure -- the high-field behavior is associated with a cluster of fermion occupation towards one corner of the plaquette, while the low-field behavior is associated with checkerboarded zero-occupation and full-occupation sites.

We are also interested in the spontaneous breakdown of chiral symmetry. 
For the $SO(3)$ QLM, a phase with spontaneously-broken chiral symmetry 
is signified by a staggered fermion occupation pattern, as well as a 
diminishing gap between the ground state and first excited state. We 
can expect a nonzero mass term to explicitly break chiral symmetry. 
To assess these properties, we examine the chiral condensate observable 
$\braket{\bar{\Psi} \Psi}$, or in our notation, the expectation value 
of $\sum_x s_x M_x$.

\begin{figure}[t!]
    \centering
    \includegraphics[width=1.0\linewidth]{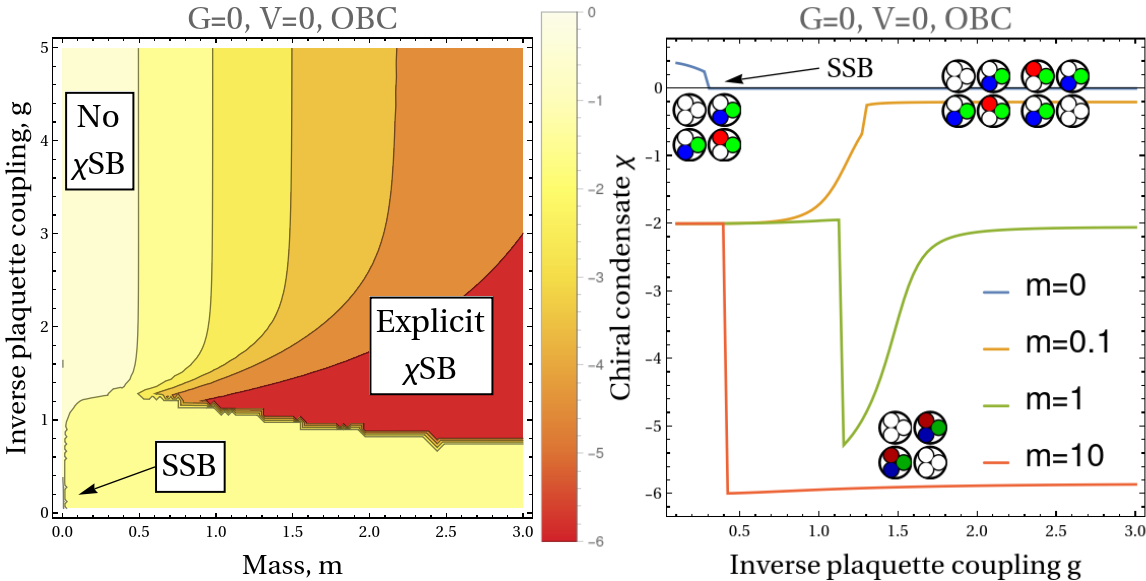}
    \caption{Ground-state chiral condensate observable $\braket{\bar{\Psi} \Psi}$ for varied fermion mass $m$ and inverse plaquette coupling $g$.}
    \label{fig:results_chiral}
\end{figure}

Via our ED on a single-plaquette, we observe how $\braket{\bar{\Psi} \Psi}$ 
varies with inverse plaquette coupling $g$ and fermion mass $m$. The 
contour plot in Figure \ref{fig:results_chiral} (left) reveals three 
distinct phases: $\braket{\bar{\Psi} \Psi} \sim 0$ (chiral symmetry preserved), 
$|\braket{\bar{\Psi} \Psi}| \sim 6$ (maximal chiral symmetry-breaking), 
and $|\braket{\bar{\Psi} \Psi}| \sim 2$ (weak chiral symmetry-breaking). 
For nonzero fermion masses, chiral symmetry-breaking is unavoidable, but 
above about $m \gtrsim 1$ the maximal and weak regions are separated by 
a sharp transition. This transition notably but imprecisely aligns with 
the low-field to high-field transition in the plaquette observable, 
as seen in Figure \ref{fig:results_magnetic}.

At $m=0$, while chiral symmetry is broadly preserved as could be 
expected, there is a critical inverse plaquette coupling $g_\chi$ 
below which chiral symmetry is again weakly broken. $g_\chi \neq g_C$, 
as more cleanly illustrated by comparing chiral condensates for 
several fixed fermion masses, seen in Figure \ref{fig:results_chiral} 
(right). Even a very small nonzero mass ($m \sim 10^{-7}$) dilutes 
this effect, returning to an explicitly symmetry-breaking regime. 
For couplings stronger than $g_\chi$ in the massless limit, the 
mass-gap is also seen to vanish.

\begin{figure}[t!]
  \centering
  \begin{minipage}[b]{0.45\textwidth}
    \includegraphics[width=\textwidth]{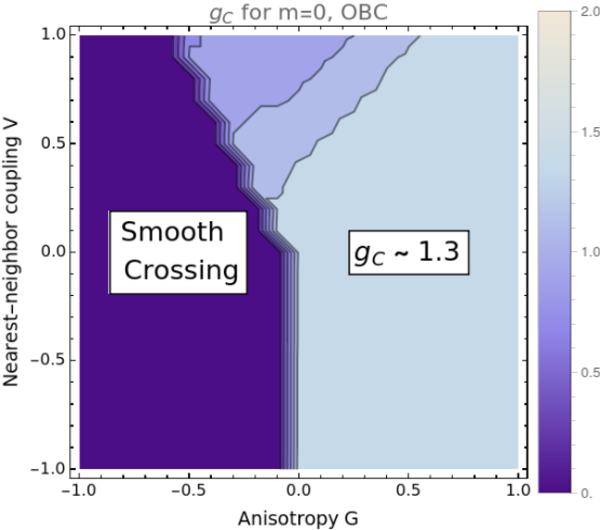}
    \caption{Magnetic critical coupling $g_C$ for massless fermions $m=0$, as a function of four-fermi couplings $G$ and $V$. In all cases the model transitions from a zero-field to a high-field regime, but to the left of the obvious boundary line this occurs smoothly, while to the right the critical value slowly retreats with increased $V$.}
    \label{fig:GV_magnetic}
  \end{minipage}
  \hfill
  \begin{minipage}[b]{0.45\textwidth}
   \includegraphics[width=\textwidth]{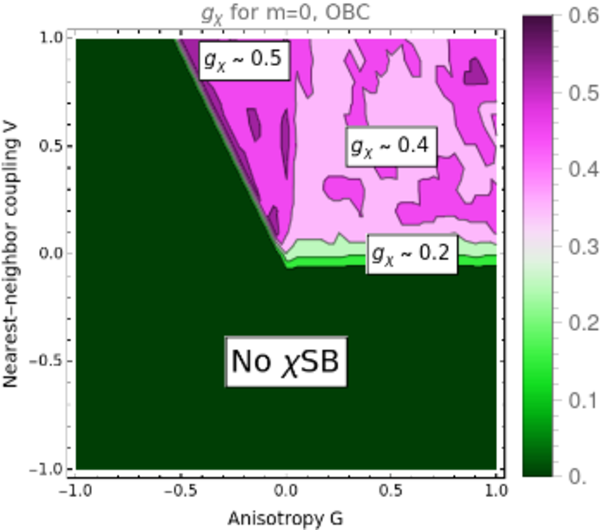}
    \caption{Spontaneous chiral symmetry-breaking critical coupling $g_\chi$ for massless fermions $m=0$, as a function of four-fermi couplings $G$ and $V$. To the left and below of the boundary line, the chiral symmetry remains unbroken, and a notably stronger plaquette coupling $g$ is required on the positive $G$-axis.}
    \label{fig:GV_chiral}
  \end{minipage}
\end{figure}

We are now in a position to ask how each of these phases is 
perturbed by the four-Fermi interactions, and 
choose to focus on the massless case. The dependence of the magnetic critical coupling $g_C$ on $G$ and $V$ is demonstrated in Figure \ref{fig:GV_magnetic}, while the same dependence of the spontaneous chiral symmetry-breaking critical coupling $g_\chi$ is seen in Figure \ref{fig:GV_chiral}.

The zero-field to  high-field transition in the plaquette observable is stable 
for positive anisotropies $G$, which energetically penalises 
sites with extremal fermion occupation ($0$ or $3$). This 
also aligns with the preliminary observation that positive 
$G$ shrinks the low-field region. For $G<0$, however, the 
zero-field to high-field transition becomes smooth, 
unless a sufficiently strong nearest-neighbor coupling $V$ is 
also introduced (numerically, the boundary is $V \sim -2G$). 
Interestingly, the critical inverse plaquette coupling $g_C$ 
only notably changes as the nearest-neighbor coupling $V$ is 
increased; it is unclear whether the transition remains 
discontinuous as $V \rightarrow \infty$, or if a critical endpoint could be identified somewhere on the $V \sim -2G$ line.

We should expect that negative nearest-neighbour coupling $V$ would preclude any chiral symmetry-breaking by favouring ``ferromagnetic'' states with $\braket{\bar{\Psi} \Psi} = 0$. This expectation is confirmed by ED, alongside a similar role of the $V \sim -2G$ diagonal line as for the magnetic observable. The spontaneous chiral symmetry-breaking phenomenon is somewhat stable for $V>0$ and $G$ to the right of this diagonal. Another notable feature is that the sign of $G$ distinguishes two regions of slightly distinct spontaneous chiral symmetry-breaking critical coupling $g_\chi$.

\section{Discussion}

While the $SO(3)$ QLM in $(2+1)d$ is an interesting quantum system in its own right, ultimately our goal is to quantum simulate QCD and thereby better understand its non-perturbative features. We should therefore take our preliminary one-plaquette results, first and foremost, as an indicator that non-Abelian QLMs with dynamical fermions are robust discretisations of gauge theories capable of exhibiting the core physics properties of QCD, including across the full phase diagram for temperature and baryo-chemical potential.

Adapting and developing quantum algorithms well-suited to simulating QLMs is an essential continuing effort, and to that end we are in the process of constructing quantum circuits to reproduce out ED results on small lattices. Because of the limited availability and reliability of quantum hardware, we are also employing large-scale classical simulations (including ED and matrix produce states), to check the stability of the single-plaquette phase diagram as the continuum is approached.

\bibliographystyle{unsrt}
\bibliography{references}

\end{document}